\documentclass{article}

\usepackage{arxiv}

\usepackage[utf8]{inputenc} 
\usepackage[T1]{fontenc}    
\usepackage{hyperref}       
\usepackage{url}            
\usepackage{booktabs}       
\usepackage{amsfonts}       
\usepackage{nicefrac}       
\usepackage{microtype}      
\usepackage{lipsum}		
\usepackage{graphicx}
\usepackage[super,sort&compress]{natbib}
\usepackage{doi}
\usepackage{amsmath,amssymb,mathtools}
\title{Structural equation modelling--based meta-analytic model for diagnostic test accuracy studies}

\author{ \href{https://orcid.org/0000-0002-3154-3855}{\includegraphics[scale=0.06]{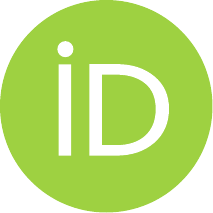}\hspace{1mm}Zelalem F. Negeri}\thanks{Corresponding author: Zelalem F. Negeri, Ph.D., Department of Statistics and Actuarial Science; University of Waterloo, M3 4116; Waterloo, Ontario, Canada N2L 3G1; Tel +1 519 888 4567, ext 4118; Email: znegeri@uwaterloo.ca.} \\
	Department of Statistics and Actuarial Science\\
	University of Waterloo\\
	Waterloo, Ontario, Canada
}

\renewcommand{\shorttitle}{SEM--based meta-analysis of DTAs}

\begin{document}
\maketitle

\begin{abstract}
Bivariate random-effects models are commonly used to jointly synthesize sensitivity and specificity in diagnostic test accuracy (DTA) meta-analysis. However, existing approaches may be sensitive to distributional assumptions and may produce biased estimates of accuracy and between-study heterogeneity with sparse meta-analytic data. We proposed a novel bivariate structural equation modelling (BSEM) approach for DTA meta-analysis and evaluated its performance relative to the bivariate linear mixed model (BLMM) and bivariate generalized linear mixed model (BGLMM). We conducted extensive simulations under varying study, sample-size, accuracy, and heterogeneity conditions. We assessed model performance using bias, root mean squared error (RMSE), confidence interval width, empirical 95\% coverage probability (CP), and convergence. We generated study-specific random effects from bivariate normal, $t$, and Laplace distributions to assess robustness to distributional misspecification. We also applied the models to two empirical DTA datasets concerning interviewer-administered questionnaires for smoking and ultrasonography for deep venous thrombosis. The BSEM generally produced smaller bias and RMSE and nominal-level CP for pooled sensitivity and specificity than the BLMM, while the BGLMM generally yielded results between the two approaches. The BLMM showed stable convergence but tended to underestimate between-study heterogeneity. We observed similar performance patterns under the three random-effects distributions. In the empirical applications, differences between model estimates were generally consistent with patterns observed in the simulations. Overall, the proposed BSEM provided a flexible alternative approach for jointly modelling sensitivity and specificity in DTA meta-analysis, with performance that was generally favourable relative to the BLMM and comparable to the BGLMM across the simulated and empirical settings considered.
\end{abstract}


\section{Introduction}
\label{S1}

Diagnostic test accuracy (DTA) studies evaluate how well a medical test discriminates between diseased and non-diseased individuals or how well a diagnostic or screening tool discriminates between a target condition and a non-target condition.\cite{Bossuyt2023DTAStudyDesign} The fundamental characteristics of DTA are sensitivity and specificity, defined as the probability of a positive test result among diseased individuals and the probability of a negative test result among non-diseased individuals, where a gold or reference standard identifies the true target condition.\cite{Bossuyt2023DTAStudyDesign, Pepe2003MedicalTests} Researchers often summarize these accuracy measures jointly using receiver operating characteristic (ROC) curves, which describe the trade-off between sensitivity and specificity across different decision thresholds.\cite{Hanley1982ROC, Hanley1983CompareROC, Walter2002SROC} DTA studies play a central role in clinical decision-making, guideline development, and health technology assessment, as they provide evidence on the usefulness of diagnostic tools.\cite{BiondiZoccai2018DiagnosticMetaAnalysis}

Meta-analysis has become an essential tool for synthesizing evidence from primary DTA studies, as individual DTA studies are often limited by small sample sizes, study-specific biases, and variability in design or study-participant populations.\cite{Borenstein2009IntroductionMetaAnalysis, Pepe2003MedicalTests} By combining results, meta-analysis increases statistical power and allows more precise estimation of test performance. Importantly, it also enables the investigation of heterogeneity, a defining feature of DTA data due to differences in thresholds, populations, and study quality.\cite{Rucker2018NetworkMetaAnalysis} 

Standard approaches to DTA meta-analysis\cite{Reitsma2005BivariateAnalysis, Chu2006BivariateMetaAnalysis} explicitly account for the correlation between sensitivity and specificity and variability within and between studies. A widely used approach is the bivariate linear mixed model (BLMM) proposed by Reitsma et al. (2005),\cite{Reitsma2005BivariateAnalysis} which describes logit-transformed sensitivity and specificity while allowing heterogeneity and correlation between studies. Closely related is the hierarchical model of Chu and Cole (2006),\cite{Chu2006BivariateMetaAnalysis} which formulates the problem using a bivariate generalized linear mixed modelling (BGLMM) approach. These models have become standard in practice\cite{Takwoingi2023UndertakingMetaAnalysis} and are implemented in major statistical software.

Despite their widespread use, standard models for DTA meta-analysis have several limitations. First, they typically assume the availability of a perfect reference standard, one that can observe true disease status without error. In real-world settings, this assumption is unrealistic, and misclassification in the reference test can bias results.\cite{Walter1999ImperfectReferenceStandards, Reitsma2009ImperfectReferenceStandard} Second, standard models rely on relatively rigid distributional assumptions,\cite{Negeri2025BivariateFiniteMixture} such as bivariate normality of random effects, and may not easily accommodate complex dependence structures, such as multiple correlated tests\cite{Steinhauser2016ModellingMultipleThresholds} or conditional dependence beyond that induced by random effects.\cite{Torrance1997DependentErrors, Dendukuri2001BayesianConditionalDependence, Dendukuri2009MultipleLatentVariableModel, Wang2017ModelingConditionalDependence} Finally, extensions to handle missing data, verification bias,\cite{Deeks2001SystematicReviewsDiagnostic, Rutjes2007EvaluationDiagnosticTests} or multiple tests\cite{Dendukuri2009MultipleLatentVariableModel} can become cumbersome within the traditional framework for DTA meta-analysis.

These limitations motivate structural equation modelling (SEM) as a flexible, unifying alternative. SEM provides a general framework for modelling latent variables, measurement error, and hierarchical structures within a single system, and its utility has been demonstrated in the context of meta-analysis of intervention studies using frequentist and Bayesian approaches.\cite{Wilson2016FittingMASEM, Cheung2016TSSEMOSMASEM, Jak2016MASEMMissingCorrelations, Mels2016OneStageMASEM, Yu2016BayesianMASEM, Cheung2013MultivariateSEM, Cheung2014MASEMR, Cheung2015MetaAnalysisSEM} In the context of DTA meta-analysis, SEM would also allow explicit modelling of latent disease status when no gold standard is available, incorporation of measurement error in both index and reference tests, and flexible representation of between-study heterogeneity and correlation structures, among others. Moreover, SEM naturally connects to latent class models, enabling principled extensions to settings with imperfect reference standards.

The primary aim of this manuscript is to introduce, for the first time to the best of the author's knowledge, a bivariate structural equation modelling (BSEM) approach for meta-analysis of diagnostic test accuracy (DTA) studies and to evaluate its performance relative to two established approaches, the BLMM\cite{Reitsma2005BivariateAnalysis} and the BGLMM.\cite{Chu2006BivariateMetaAnalysis} The proposed BSEM jointly models sensitivity and specificity while accounting for between-study variation and their covariance. We evaluate its performance through an extensive simulation study across different levels of diagnostic accuracy, numbers of studies, within-study sample sizes, and random-effects distributions, and through applications to two empirical DTA datasets. The manuscript also provides an implementation of the proposed approach using the \texttt{R}\cite{RCoreTeam2025R} statistical computing language and the \texttt{OpenMx}\cite{Boker2025OpenMx, Neale2016OpenMx2, Pritikin2015OpenSourceIFA, Hunter2018StateSpaceModeling} and \texttt{metaSEM}\cite{Cheung2015metaSEM} packages to facilitate its application by researchers and practitioners. Finally, potential extensions of the BSEM framework to more complex DTA settings, including imperfect reference standards and multiple diagnostic tests, are discussed as directions for future research.

The remainder of this manuscript is organized as follows. Section \ref{S2} presents two empirical data examples to motivate the need for SEM-based meta-analytic models. Section \ref{S3} provides an overview of the standard methods and describes our new model in detail. In Section \ref{S4}, we present the methods' finite-sample properties through an extensive simulation study, followed by an illustration with real-world data examples in Section \ref{S5}. We close the paper with a discussion and conclusion in Section \ref{S6}.

\section{Motivating examples}
\label{S2}
In this section, we present two real-life DTA meta-analytic datasets that we analyze using the traditional and proposed approaches later in Section \ref{S5}.

\subsection{IAQ data}
The validity of self-reported smoking was investigated by Patrick \textit{et al.} (1994)\cite{Patrick1994ValiditySelfReportedSmoking} using biochemical measures of smoking as reference standards. The review included 26 published reports comprising 51 comparisons between self-reported smoking and biochemical assessments. The \texttt{mada}\cite{Doebler2012Mada} package provides the interviewer-administered questionnaire (IAQ) data as a subset of these studies for assessing smoking status. The data (see Appendix Table A.1) are presented as $2\times 2$ tables containing true positives, false negatives, false positives, and true negatives, with some studies contributing more than one table.

\begin{center}
    \begin{figure}[ht]
        \centering
        \includegraphics[width = \textwidth]{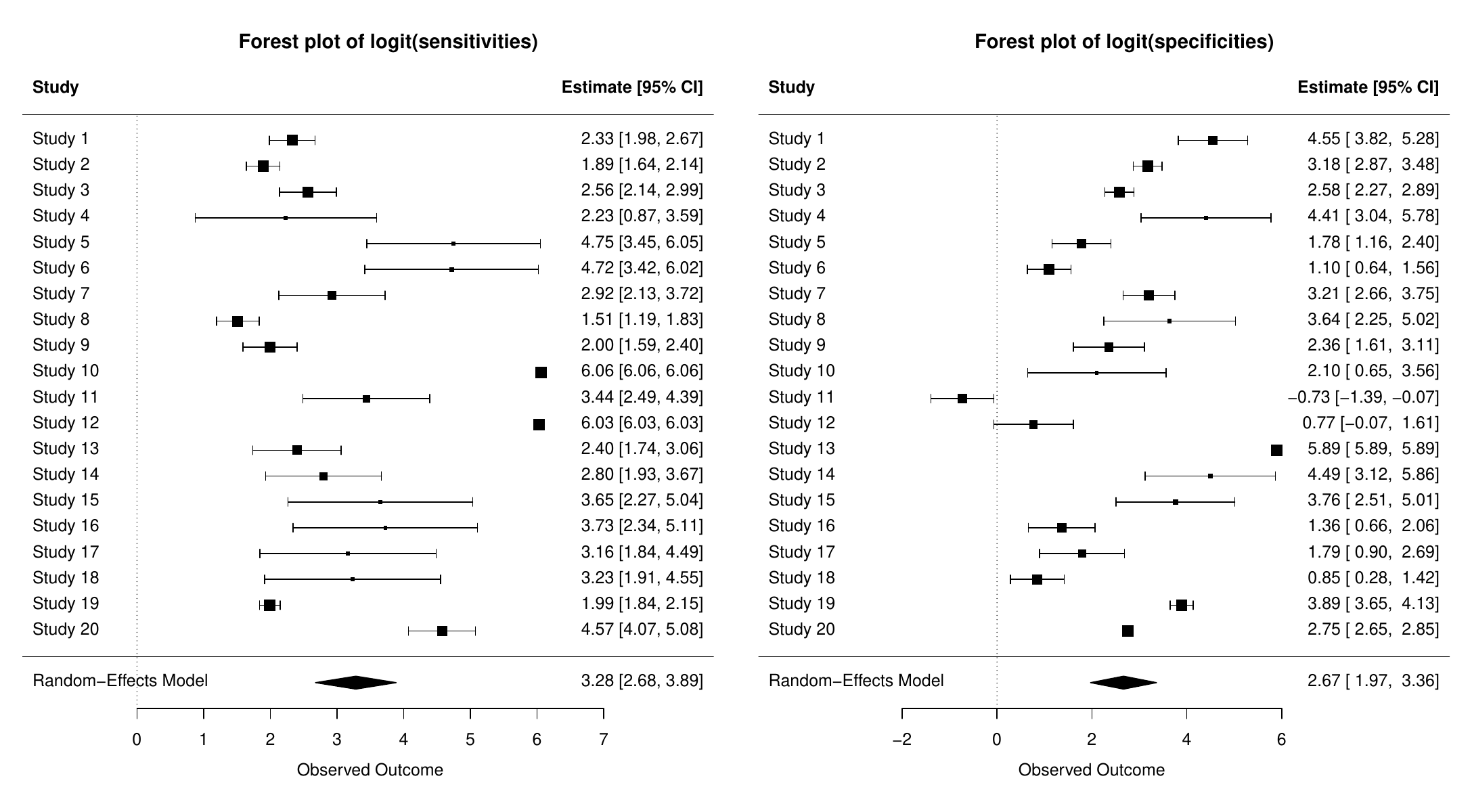}
        \caption{Forest plot of logit-sensitivity (left) and logit-specificity (right) for the IAQ data (Patrick \it{et al}., 1994)\cite{Patrick1994ValiditySelfReportedSmoking}}
        \label{Fig1}
    \end{figure}
\end{center}

Figure \ref{Fig1} presents the forest plot of sensitivity and specificity, on the logit scale, for the IAQ data. The plot includes 20 study-specific estimates and shows substantial heterogeneity in both sensitivity and specificity across studies. On average, the IAQ data contained about 350 smoker participants and about 730 non-smoker participants. Additionally, Studies 10 and 12 appear outlying in sensitivity, and Study 11 in specificity. Several studies show relatively wide confidence intervals (CIs), whereas other studies have estimates close to the upper boundary with very narrow CIs. The variation in study-specific diagnostic accuracy and precision makes these data suitable for illustrating the proposed BSEM approach.

\subsection{DVT data}
Kearon \textit{et al.} (1998)\cite{Kearon1998NoninvasiveDiagnosis} investigated the accuracy of venous ultrasonography for diagnosing deep venous thrombosis (DVT). The \texttt{metadat}\cite{Viechtbauer2025metadat} package provides these data, which include diagnostic accuracy results from studies evaluating venous ultrasonography for diagnosing a first episode of DVT in symptomatic and asymptomatic patients. Contrast venography determined the presence or absence of DVT, and venous ultrasonography was evaluated as the diagnostic test. The dataset contains study-level information, including patient status (asymptomatic vs symptomatic) and the numbers of true positives, positive patients, true negatives, and negative patients, from which the corresponding $2\times 2$ diagnostic tables can be obtained (see Appendix Table A.2).

Figure \ref{Fig2} presents the forest plot of sensitivity and specificity for the 30 studies included in the analysis, with both measures presented on the logit scale. Although not as high as for the IAQ dataset, considerable variation is observed in the study-specific specificity estimates and their precision. The review contains an average of about 50 and 96 participants with and without DVT, respectively. In addition, several studies have very narrow CIs, whereas others have considerably wider CIs. These characteristics provide substantial between-study heterogeneity in the observed specificity and make the DVT data suitable for illustrating the application of our proposed approach in Section \ref{S5}.

\begin{center}
    \begin{figure}[ht]
        \centering
        \includegraphics[width = \textwidth]{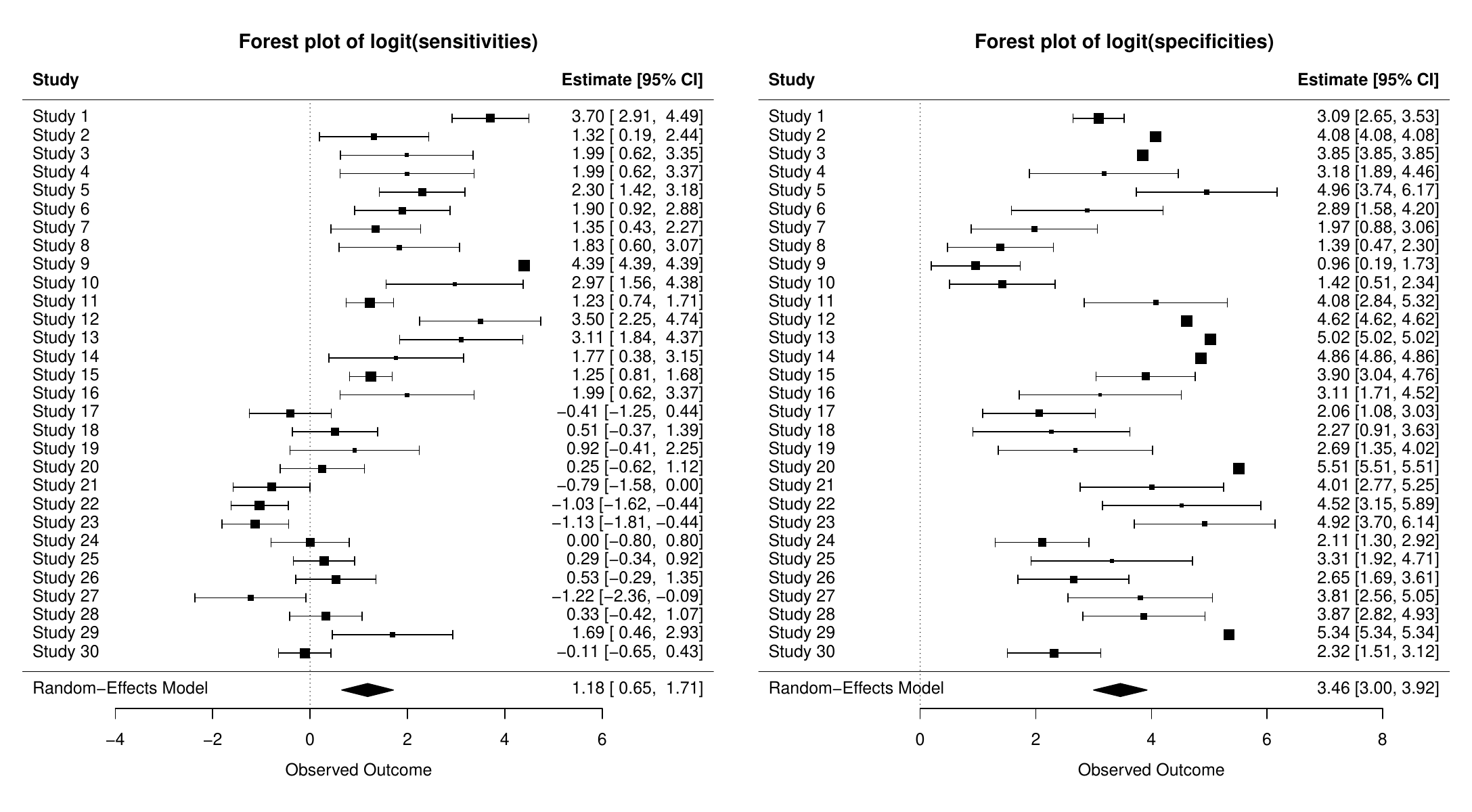}
        \caption{Forest plot of logit-sensitivity (left) and logit-specificity (right) for the DVT data (Kearon \it{et al}., 1998)\cite{Kearon1998NoninvasiveDiagnosis}}
        \label{Fig2}
    \end{figure}
\end{center}

\section{Methods}
\label{S3}
This section presents the statistical methods for two of the most widely used approaches to aggregate-data meta-analysis (ADMA) of DTA studies, BLMM\cite{Reitsma2005BivariateAnalysis} and BGLMM,\cite{Chu2006BivariateMetaAnalysis} as well as our newly proposed BSEM approach.

\subsection{Standard models for DTA meta-analysis}
Assuming $k$ studies in the meta-analysis, indexed by $i=1,\cdots,k,$ the BLMM of Reitsma \textit{et al.} (2005)\cite{Reitsma2005BivariateAnalysis} incorporates the within-study variation in test sensitivity and specificity using an approximate bivariate normal distribution as

\begin{equation}
	\label{E1}
	\left[\begin{array}{c}
	g(\widehat{Se}_{i})\\
	g(\widehat{Sp}_{i})
	\end{array} \right] \; \dot\sim \; N_2\left[\left(\begin{array}{c}
	\phi_{1i}\\
	\phi_{2i}
	\end{array}\right),\bf{S}_i = \left(\begin{array}{cc}
	s^2_{1i} & 0\\
	0 & s^2_{2i}
	\end{array}\right)\right], \; \; i=1,...,k,
\end{equation}
where g$(\cdotp)$ denotes a known transformation and $\textbf{S}_i$ represents the within-study covariance matrix for [g$(\widehat{Se}_{i})$, g$(\widehat{Sp}_{i})]^T$.

The BLMM further assumes a bivariate normal distribution for the random effects that capture the heterogeneity between studies
\begin{equation}
	\label{E2}
	\left(\begin{array}{c}
	\phi_{1i}\\
	\phi_{2i}
	\end{array} \right) \; \dot\sim \; N_2\left[{\pmb{\mu}}=\left(\begin{array}{c}
	\mu_1\\
	\mu_2
	\end{array}\right),\pmb{\Sigma}= \left(\begin{array}{cc}
	\sigma^2_1 & \sigma_{12}\\
	\sigma_{12} & \sigma^2_2
	\end{array}\right)\right], \; \; i=1,...,k.
\end{equation}
The marginal model of the BLMM is completed by combining \eqref{E1} \& \eqref{E2}:
\begin{equation}
	\left[\begin{array}{c}
	g(\widehat{Se}_{i})\\
	g(\widehat{Sp}_{i})
	\end{array} \right] \; \dot\sim \; N_2\left[{\pmb{\mu}}=\left(\begin{array}{c}
	\mu_1\\
	\mu_2
	\end{array}\right),\pmb{\Sigma}_i = \left(\begin{array}{cc}
	\sigma^2_1+s^2_{1i} & \sigma_{12}\\
	\sigma_{12} & \sigma^2_2+s^2_{2i}
	\end{array}\right)\right]. \nonumber
\end{equation}
The within-study heterogeneity variances are calculated from the data and used to enter the likelihood contribution of each study as known data characteristics. Several authors have suggested either asymptotic or exact approaches to calculating these variances. For example,
\begin{itemize}
    \item For the BLMM-Logit (Reitsma \textit{et al.,} 2005),\cite{Reitsma2005BivariateAnalysis} the variances are approximated as:
\begin{align*}
	\text{g}(\hat{p}_i) &= \text{logit}(\hat{p}_i), i=1,...,k, \; \text{where} \; \hat{p}_i = \widehat{Se}_i \; \text{or} \; \widehat{Sp}_i\\
	s^2_{ji} &\approx \frac{1}{n_{ji}\hat{p}_i(1-\hat{p}_i)}, \; j=1,2.
\end{align*}
    \item For the arcsine square root transformation (BLMM-ASR; Negeri \textit{et al.,} 2018),\cite{Negeri2018BivariateRandomEffects} the approximate variances are:
\begin{align*}
	\text{g}(\hat{p}_i) & = \text{sin}^{-1}\left(\sqrt{\hat{p}_i}\right), i=1,...k,\\
	s^2_{ji} &\approx \frac{1}{4n_{ji}}, \; j=1,2.
\end{align*}
	\item For the Freeman-Tukey double arcsine transformation (BLMM-FTDA; Negeri \textit{et al.,} 2018),\cite{Negeri2018BivariateRandomEffects} the approximate variances are:
\begin{align*} 
	\text{g}(\hat{p}_{i}) & = (1/2)\left[ \hspace{0.1cm}\text{sin}^{-1}\left(\sqrt{\frac{n_{ji}\hat{p}_i}{n_{ji}+1}}\right) + \text{sin}^{-1}\left(\sqrt{\frac{n_{ji}\hat{p}_i + 1}{n_{ji}+1}}\right)\right], \\
	s^2_{ji} &\approx \frac{1}{4n_{ji}+2}, \; j=1,2.
\end{align*}
    \item For the BLMM-Exact (Dabi and Negeri, 2026),\cite{Dabi2026ExactWithinStudyVariance} the exact variances, irrespective of the transformation used, are given by:
\begin{align*} 
	s_{ji}^2 
	& = \sum_{x=0}^{n_{ji}} \Bigg[g\!\left(\tfrac{x}{n_{ji}}\right)\Bigg]^2 
	\binom{n_{ji}}{x} p_{ji}^x (1-p_{ji})^{\,n_{ji}-x} \\
	&\quad - \left[\sum_{x=0}^{n_{ji}} g\!\left(\tfrac{x}{n_{ji}}\right) 
	\binom{n_{ji}}{x} p_{ji}^x (1-p_{ji})^{\,n_{ji}-x}\right]^2, \; j=1,2.
\end{align*}	
\end{itemize}
The parameters of the BLMM are estimated iteratively using the maximum likelihood (ML) method. For the primary parameters of interest $(\boldsymbol{\mu})$, the ML estimator is equivalent to the generalized re-weighted least squares solutions:
\begin{equation}
	\widehat{\boldsymbol{\mu}} = \left[\sum_{i=1}^{k}(\widehat{\boldsymbol{\Sigma}}+\textbf{S}_i)^{-1}\right]^{-1}\left[(\widehat{\boldsymbol{\Sigma}}+\textbf{S}_i)^{-1}\textbf{y}_i \right], \label{E2.10}
\end{equation}
where $\textbf{y}_i$=$[$g$(\widehat{Se}_{i}),\; $g$(\widehat{Sp}_{i})]'$. Thus, the estimated variance-covariance matrix of the parameter of interest estimator is given by
\begin{equation}
	\widehat{\text{Var}(\boldsymbol{\widehat{\mu}})} = \left[\sum_{i=1}^{k}(\widehat{\boldsymbol{\Sigma}}+\textbf{S}_i)^{-1}\right]^{-1} \label{E2.12}.
\end{equation}
Inferences, such as construction of CIs for ${\boldsymbol{\mu}}$ can be carried out based on asymptotic theory and utilizing equations \ref{E2.10} and \ref{E2.12}.

The BGLMM of Chu and Cole (2006),\cite{Chu2006BivariateMetaAnalysis} which is considered the preferred approach for ADMA of DTAs, and models the within-study variability in test sensitivity and specificity using the Binomial distribution:
\begin{align*}
	TP_i|Se_i & \sim \text{Binomial}(n_{1i}, Se_i); \hspace{0.1cm} \text{g}(Se_i) = \mu_1 + \phi_{1i}, \; i=1,\cdots, k\\
	TN_i|Sp_i & \sim \text{Binomial}(n_{2i}, Sp_i); \hspace{0.1cm} \text{g}(Sp_i) = \mu_2 + \phi_{2i}, \; i=1,\cdots, k
\end{align*}
As in the BLMM, the bivariate normal distribution is commonly assumed to model the between-study variability in test sensitivity and specificity as:
\begin{align}
	\left(\phi_{1i}, \phi_{2i}\right)^T & \sim N_2(\mathbf{0}, \boldsymbol{\Sigma}), \nonumber
\end{align}
where g$(\cdotp)$ is commonly the logit link function. Adaptive Gauss-Hermite quadrature (AGHQ) is commonly used to approximate the likelihood function, followed by numerical optimization of the approximation to iteratively obtain ML estimators of the BGLMM parameters.

\subsection{Proposed approach for DTA meta-analysis}
In this subsection, we present the proposed BSEM formulation of standard models for ADMA of DTA studies. BSEM provides an alternative but equivalent formulation of the standard BLMM.\cite{Reitsma2005BivariateAnalysis} However, the BSEM framework allows the model to be expressed in terms of latent variables and path structures, which facilitates both interpretability and extension to more complex situations.

As defined earlier, let $\textbf{y}_i$=$[$g$(\widehat{Se}_{i}),\; $g$(\widehat{Sp}_{i})]'$ denote the observed logit-transformed sensitivity and specificity for study $i$. BSEM's measurement model is equivalent to BLMM's within-study model and can be expressed in SEM notation as:

\begin{equation}
\textbf{y}_i= \boldsymbol{\phi}_i + \boldsymbol{\varepsilon}_i, \; \boldsymbol{\varepsilon}_i\; \sim \; N_2(\mathbf{0}, \textbf{S}_i),
\label{SEM_1}
\end{equation}
where $\boldsymbol{\phi}_i=(\phi_{1i}, \phi_{2i})'$ are the latent variables, and $\textbf{S}_i$ captures the measurement error.

BSEM's structural model is also equivalent to the between-study model of the BLMM that describes the heterogeneity across studies:
\begin{equation} 
\boldsymbol{\phi}_i \; \sim \; N_2(\boldsymbol{\mu}, \boldsymbol{\Sigma}),
\label{SEM_2}
\end{equation}
where the latent variables $\boldsymbol{\varepsilon}_i$ vary across studies and their covariance structure $\boldsymbol{\Sigma}$ represent random effects.

Finally, the combined BSEM representation is obtained by substituting the structural model \eqref{SEM_1} into the measurement model \eqref{SEM_2} as:
\[\textbf{y}_i = \boldsymbol{\mu} + \boldsymbol{\phi}_i + \boldsymbol{\varepsilon}_i,\]
where $\boldsymbol{\phi}_i \sim N_2(\mathbf{0}, \boldsymbol{\Sigma})$ denote the between-study random effects and $\boldsymbol{\varepsilon}_i \sim N_2(\mathbf{0}, \textbf{S}_i)$ denote the within-study error. Therefore, the BLMM and BSEM representations have the same marginal model:
\[\textbf{y}_i \sim N_2(\boldsymbol{\mu}, \boldsymbol{\Sigma}+\textbf{S}_i),\]
which demonstrates that the standard BLMM\cite{Reitsma2005BivariateAnalysis} is equivalent to a two-factor SEM with correlated latent variables and known measurement error variances.

\begin{figure}[h]
    \centering
    \includegraphics[width=\textwidth]{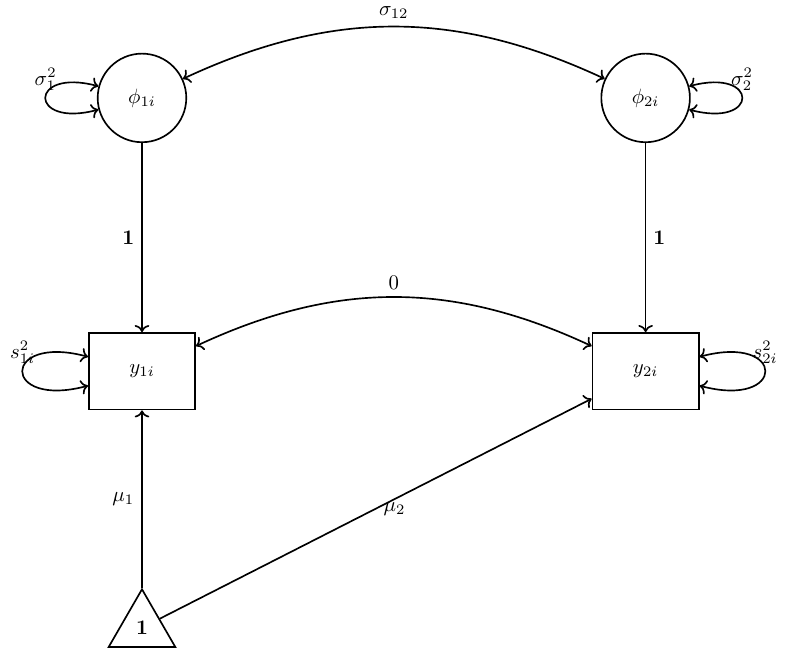}
    \caption{Path diagram depicting the proposed BSEM approach for ADMA of DTA studies}
    \label{Fig3}
\end{figure}

In addition to the mathematical description above, Figure \ref{Fig3} shows the BSEM path diagram for ADMA in DTA studies, depicting a just-identified SEM with known measurement error, which distinguishes it from typical SEM applications where error variances are estimated from the model.

This path diagram includes two latent nodes $(\phi_{1i},\phi_{2i})'$, two observed nodes $[y_{1i}=g(\widehat{Se}_i),y_{2i}=g(\widehat{Sp}_i)]'$, two directed arrows from latent variables to observed variables with factor loadings fixed at 1, a double-headed arrow between latent variables denoting their covariance $(\sigma_{12})$, and residual variances fixed to $s_{1i}^2$ and $s_{2i}^2$.

By using definition variables to impose the known within-study covariance matrices, $\textbf{S}_i$, one can maximize the following log-likelihood function via the full information maximum likelihood (FIML) approach (Enders, 2010)\cite{Enders2010AppliedMissingData} to obtain ML estimators of the proposed BSEM approach:
\begin{align*}
	\ell(\boldsymbol{\theta}; \mathbf{y}_i) &= -\frac{k}{2} \log(2\pi) - \frac{1}{2} \sum_{i=1}^{k} \log|\boldsymbol{\Sigma}_i(\boldsymbol{\theta})|  - \frac{1}{2} \sum_{i=1}^{k} (\mathbf{y}_i - \boldsymbol{\mu}(\boldsymbol{\theta}))^T \boldsymbol{\Sigma}_i^{-1} (\mathbf{y}_i - \boldsymbol{\mu}(\boldsymbol{\theta})),
\end{align*}
by setting $\boldsymbol{\mu}_i(\boldsymbol{\theta}) = \boldsymbol{\mu}$ and $\boldsymbol{\Sigma}_i(\boldsymbol{\theta}) = \boldsymbol{\Sigma}_i = \boldsymbol{\Sigma} + \mathbf{S}_i.$

The proposed BSEM representation enables seamless inference, including hypothesis testing and construction of CIs for either the fixed-effects or random-effects parameters, using asymptotic theory for ML estimators. See Section \ref{S6} for an extensive discussion of how the proposed BSEM can easily be extended to handle other DTA meta-analysis settings, including imperfect reference standards, multiple diagnostic tests, missing data, and verification bias.
        
\subsection{Model implementations in \texttt{R}}
We have successfully implemented the proposed meta-analytic BSEM approach for ADMA in DTA studies by adapting functions in the \texttt{metaSEM}\cite{Cheung2015metaSEM} and \texttt{OpenMX}\cite{Neale2016OpenMx2} \texttt{R}\cite{RCoreTeam2025R} packages. The \texttt{metaSEM}\cite{Cheung2015metaSEM} package implements a two-stage meta-analytic SEM approach for meta-analysis of intervention studies in which covariance matrices are pooled in stage 1, and SEM is fit in stage 2. The package uses a simpler syntax and enables faster implementation. However, if one needs more flexibility to accommodate custom likelihoods that implement imperfect reference standards, for example, the \texttt{OpenMx}\cite{Neale2016OpenMx2} package can be used.

Following Cheung (2015),\cite{Cheung2015MetaAnalysisSEM} we fitted the proposed BSEM approach using ML, tested the equality of heterogeneity variances across studies using the likelihood ratio (LR) test statistic, and constructed 95\% confidence ellipses around the observed and pooled logit-transformed sensitivity and specificity. In addition, we employed the Huber-White robust ``sandwich estimator"\cite{Huber1967MaximumLikelihood, White1980Heteroskedasticity, Arminger1989PseudoMaximumLikelihood, Savalei2014RobustCorrections} for the BSEM to obtain robust standard errors to mitigate the effect of outlying studies on CIs. We fitted the standard BLMM\cite{Reitsma2005BivariateAnalysis} and the BGLMM\cite{Chu2006BivariateMetaAnalysis} using the Laplace approximation algorithm with the \texttt{mada}\cite{Doebler2012Mada} and \texttt{lme4}\cite{Bates2015Lme4} packages in \texttt{R},\cite{RCoreTeam2025R} respectively. For all data analysis, we used \texttt{R}\cite{RCoreTeam2025R} version 4.5.2 (2025-10-31) and \texttt{RStudio}\cite{Posit2026RStudio} version 2026.8.2.200.

\section{Simulation study}
\label{S4}

\subsection{Simulation study design}
We conducted an extensive simulation study to evaluate the finite-sample properties of the proposed BSEM approach compared with traditional models for ADMA of DTA studies. We varied the true model parameters $(\mu_1, \mu_2, \sigma_1^2, \sigma_2^2, \sigma_{12})$ and data characteristics, such as the number of studies and participants within a study, informed by the characteristics of the two real-life datasets presented in Section \ref{S2} and the simulation design of Dabi and Negeri.\cite{Dabi2026ExactWithinStudyVariance} The true population parameters used to generate DTA meta-analytic data are presented in Table \ref{Tab1}. In total, we considered $3^3\times 5 = 135$ scenarios and generated random effects from three bivariate distributions: normal, $t$ with three degrees of freedom, and Laplace to evaluate robustness of the models to distributional misspecifications. 

\begin{table}[ht]
\centering
\caption{True model parameter settings for the simulation study}
\label{Tab1}
\begin{tabular}{@{}lccccc@{}}
\toprule
\textbf{Parameter} & \textbf{Setting 1} & \textbf{Setting 2} & \textbf{Setting 3} & \textbf{Setting 4} & \textbf{Setting 5}\\
\midrule
(Se, Sp) & (0.70, 0.80) & (0.80, 0.90) & (0.95, 0.99) & $-$ & $-$\\
($\sigma_1^2, \sigma_2^2$) & (0.40, 0.40) & (0.80, 0.80) &  (1.20, 3.00) & $-$ & $-$\\
$\sigma_{12}$ & -0.30 & -0.50 & -0.80 & $-$ & $-$\\
($n_1, n_2$) & (40, 50) & (90, 200) & (300, 700) & $-$ & $-$\\
$k$ & 5 & 20 & 40 & 60 & 100\\
\bottomrule
\end{tabular}
\end{table}

For each simulation replicate, we generated $k$ diagnostic accuracy studies comprising the meta-analytic data. For study $i$, we simulated the study-specific sensitivity and specificity on the logit scale $(\boldsymbol{\phi}_i)$ with mean vector $\boldsymbol{\mu}$ and between-study covariance $\boldsymbol{\Sigma}$. Then we back-transformed the generated logit-scale values using the inverse-logit function\cite{faraway2025} to obtain the study-specific sensitivity ($Se_i$) and specificity ($Sp_i$). Conditional on $Se_i$ and $Sp_i$, we generated the numbers of true positives and true negatives independently as $TP_i\sim \text{Binomial}(n_{1i}, Se_i)$ and $TN_i\sim \text{Binomial}(n_{2i}, Sp_i)$, where $n_{1i}$ and $n_{2i}$ denote the numbers of diseased and non-diseased participants in study $i$, respectively. We then calculated the numbers of false negatives and false positives as $FN_i=n_{1i}-TP_i$ and $FP_i=n_{2i}-TN_i$, respectively. The four resulting cell counts constituted the $2\times2$ diagnostic table for each study. We repeated the process for the specified combinations of true sensitivity and specificity, between-study variance and covariance parameters, numbers of studies, and within-study sample sizes. We employed the \texttt{mvtnorm}\cite{mvtnorm2026} package to simulate data from the bivariate normal and $t$ distributions and \texttt{LaplacesDemon}\cite{LaplacesDemon2026} to generate data from the bivariate Laplace distribution.

We replicated each simulation scenario $1,000$ times to evaluate the finite-sample performance of the three competing models. For each scenario, we assessed model performance using bias, root mean squared error (RMSE), empirical 95\% coverage probability (CP), and 95\% CI width.\cite{Burton2006Simulation, Morris2019Simulation} To ensure reproducibility, we used replication-specific unique seed numbers in \texttt{R}.\cite{RCoreTeam2025R} We assessed convergence of the BGLMM and BSEM models using the convergence indicators provided by the \texttt{lme4}\cite{Bates2015Lme4} and \texttt{metaSEM}\cite{Cheung2015metaSEM} packages, respectively, specifically the \texttt{\$optinfo\$conv\$lme4\$messages} component for BGLMM and \texttt{\$Mx.status1} for BSEM.

\subsection{Simulation results}
\label{S4.1}
This section presents simulation results for different combinations of the true model parameters and data characteristics. Although we primarily discuss small-sample settings to evaluate the models' robustness in extreme cases, results for other scenarios are available in the appendices of the online \textit{Ancillary files}. We also present results based on normally distributed random effects as our primary findings and discuss the effects of other random-effects distributions.

\begin{center}
    \begin{figure}[h]
        \centering
        \includegraphics[width = \textwidth]{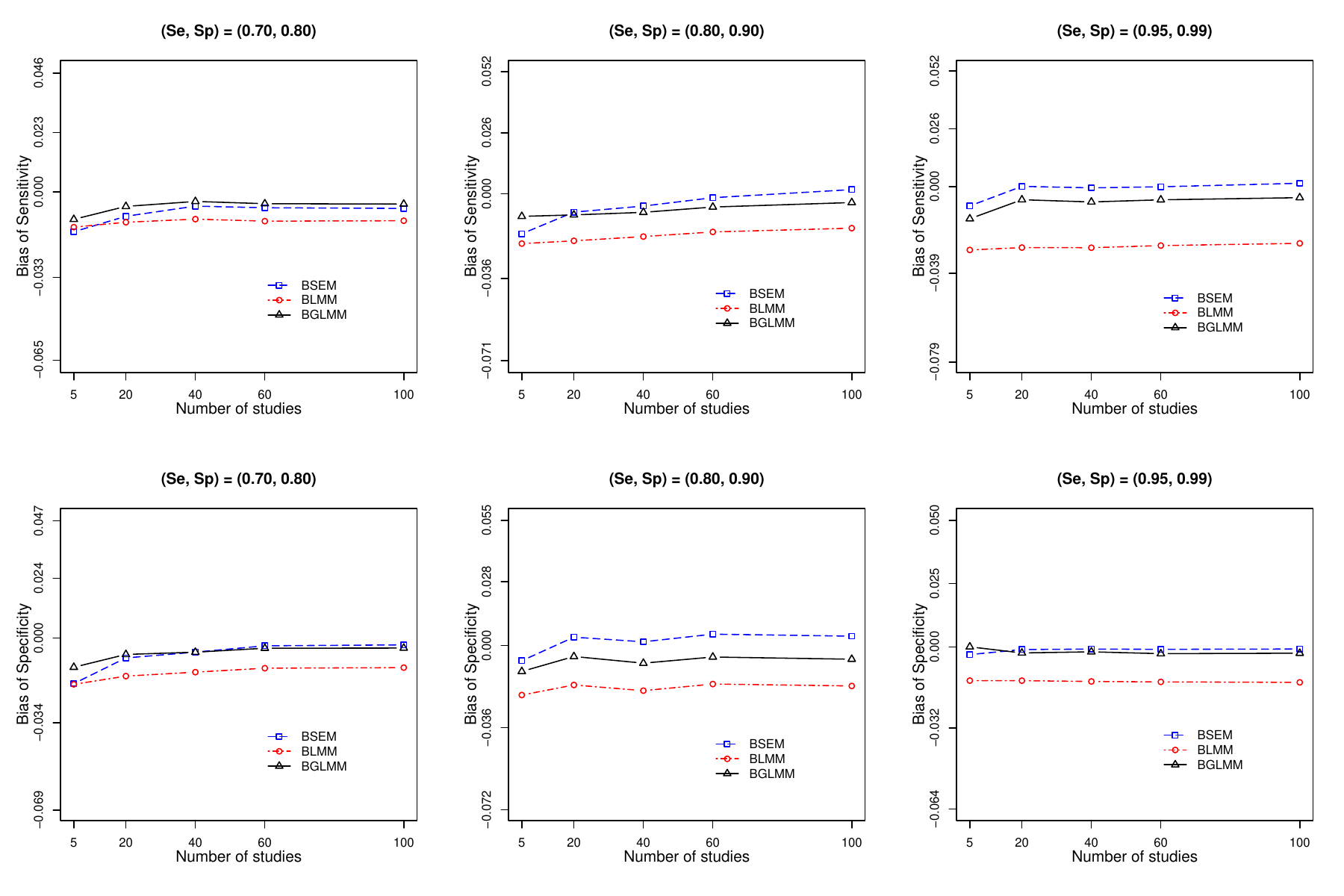}
        \caption{Bias of pooled sensitivity and specificity for different combinations of true sensitivity, true specificity, and number of studies when $\sigma_1^2=\sigma_2^2=0.80$, $\sigma_{12}=-0.50$, $n_1=40$, $n_2=50$ and random effects were generated from the bivariate normal distribution.}
        \label{Fig4}
    \end{figure}
\end{center}

Regarding the bias of the pooled sensitivity and specificity estimators (Figure \ref{Fig4}), the three models underestimated both parameters in the low $(Se=0.70, Sp=0.80)$ or moderate $(Se=0.80, Sp=0.90)$ diagnostic accuracy settings. In particular, the standard BLMM's negative bias was dominant across the three diagnostic accuracy levels, whereas the proposed BSEM generally yielded the least biased estimates. The standard BGLMM approach produced bias values between those of the BLMM and the BSEM, especially in moderate $(Se=0.80, Sp=0.90)$ to high $(Se=0.95, Sp=0.99)$ diagnostic accuracy settings. Moreover, although the biases of the BSEM and BGLMM improved with increasing numbers of studies, those of the BLMM remained constant. However, all three models yielded unbiased and almost indistinguishable results when the number of participants in the diseased and non-diseased groups increased to $n_1=300$ and $n_2=700$, respectively, irrespective of the values of the other parameters (Appendix Figures B.3, B.6 and B.9). The bias of the three models was not substantially affected by the assumed distribution of the random effects, as these patterns remained consistent when the random effects were generated from the bivariate t and Laplace distributions (Appendix Figures B.10 to B.27).

\begin{center}
    \begin{figure}[h]
        \centering
        \includegraphics[width = \textwidth]{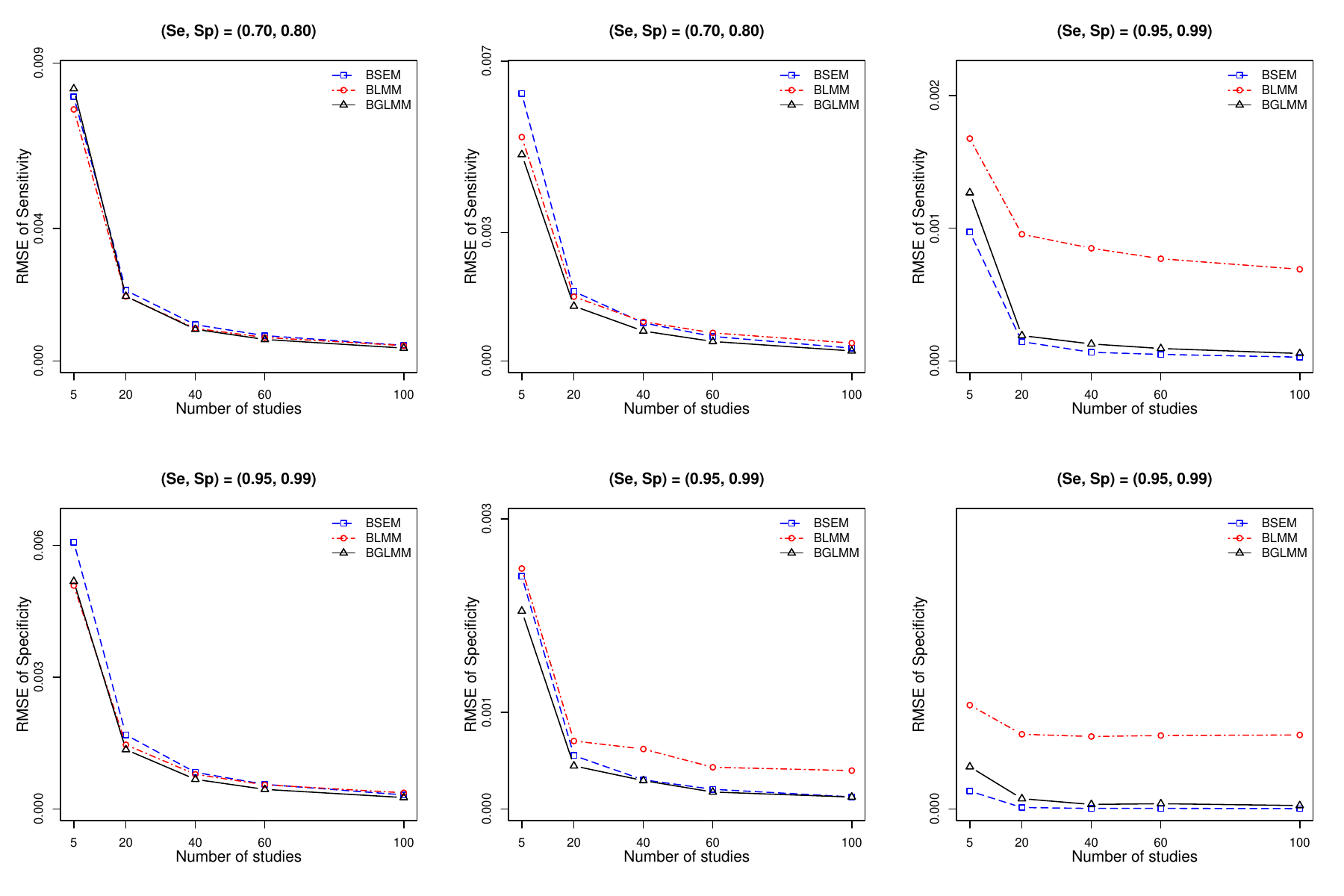}
        \caption{RMSE of pooled sensitivity and specificity for different combinations of true sensitivity, true specificity, and number of studies when $\sigma_1^2=\sigma_2^2=0.80$, $\sigma_{12}=-0.50$, $n_1=40$, $n_2=50$ and random effects were generated from the bivariate normal distribution.}
        \label{Fig5}
    \end{figure}
\end{center}

Figure \ref{Fig5} shows the RMSE of the pooled sensitivity and specificity estimators for the same simulation scenarios as the bias figure. As expected, the RMSE of all three models decreased with increasing numbers of studies. Although the models had comparable RMSE values for low-to-moderate-diagnostic accuracy settings, standard BLMM had higher RMSE values than BGLMM and BSEM for high-diagnostic accuracy settings. However, in scenarios with large sample sizes within a study, the models had indistinguishable, near-zero RMSE values, especially as the number of studies increased, regardless of the true values of other parameters (Appendix Figures C.3, C.6, and C.9). The results were also consistent across the three distributions used to generate the random effects, with no substantial change in the relative performance of the three models (Appendix Figures C.10 to C.27).

\begin{center}
    \begin{figure}[h]
        \centering
        \includegraphics[width = \textwidth]{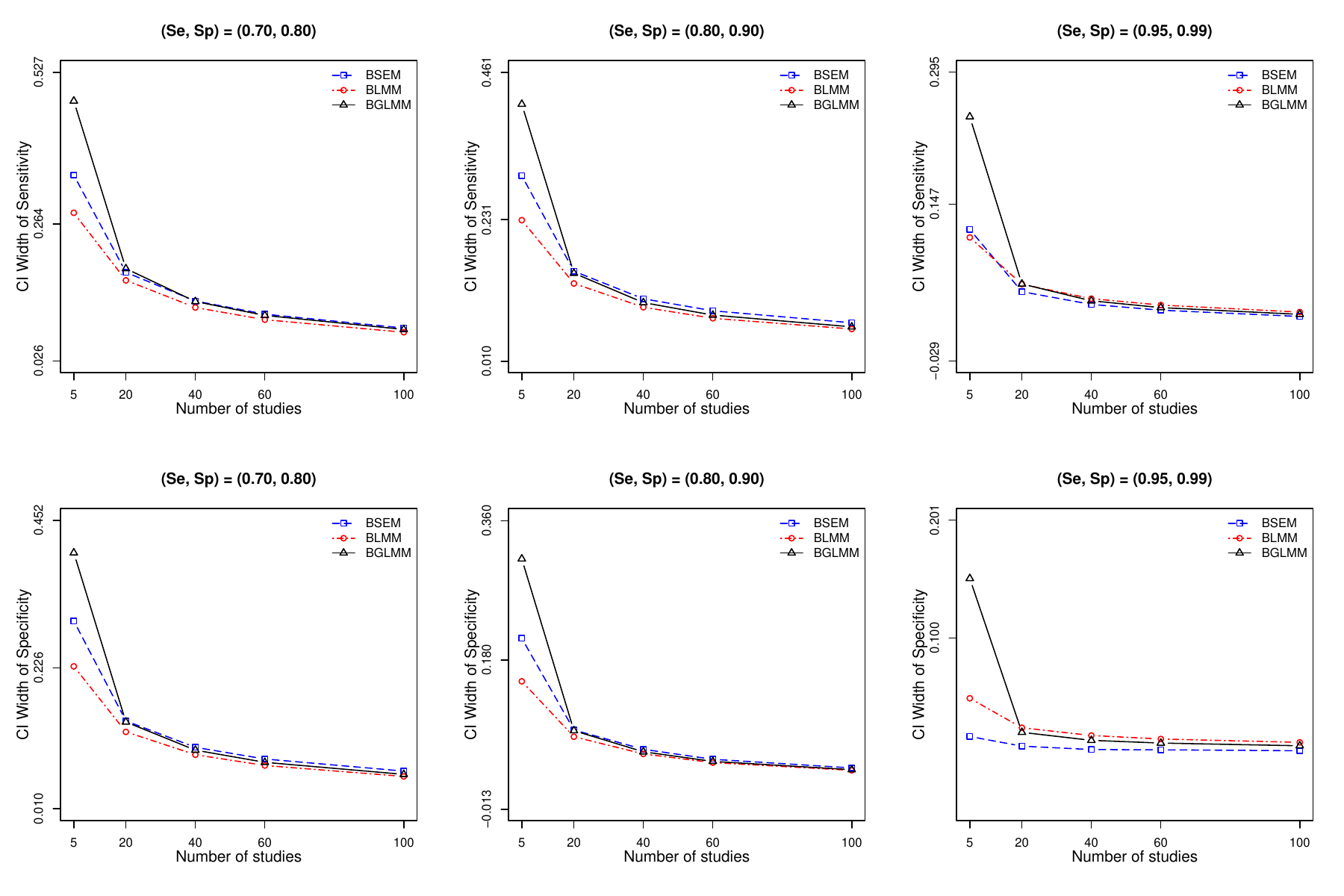}
        \caption{Confidence interval widths of pooled sensitivity and specificity for different combinations of true sensitivity, true specificity, and number of studies when $\sigma_1^2=\sigma_2^2=0.80$, $\sigma_{12}=-0.50$, $n_1=40$, $n_2=50$ and random effects were generated from the bivariate normal distribution.}
        \label{Fig6}
    \end{figure}
\end{center}

Figure \ref{Fig6} presents the average 95\% CI widths for the true sensitivity and specificity of the three models when random effects were generated from the bivariate normal distribution and sample sizes within the study were small. Regardless of the level of diagnostic accuracy, the three models produced CI widths that decreased with increasing meta-analysis size. Moreover, the models had comparable CI widths for larger meta-analyses $(k\geq 20)$. However, the standard BGLMM had wider CIs, and the BLMM and BSEM had narrower CIs for meta-analyses with fewer than $20$ studies, regardless of the true diagnostic accuracy values. These results were consistent across other parameter combinations and random-effects distributions (Appendix Figures D.1 to D.27).

\begin{center}
    \begin{figure}[h]
        \centering
        \includegraphics[width = \textwidth]{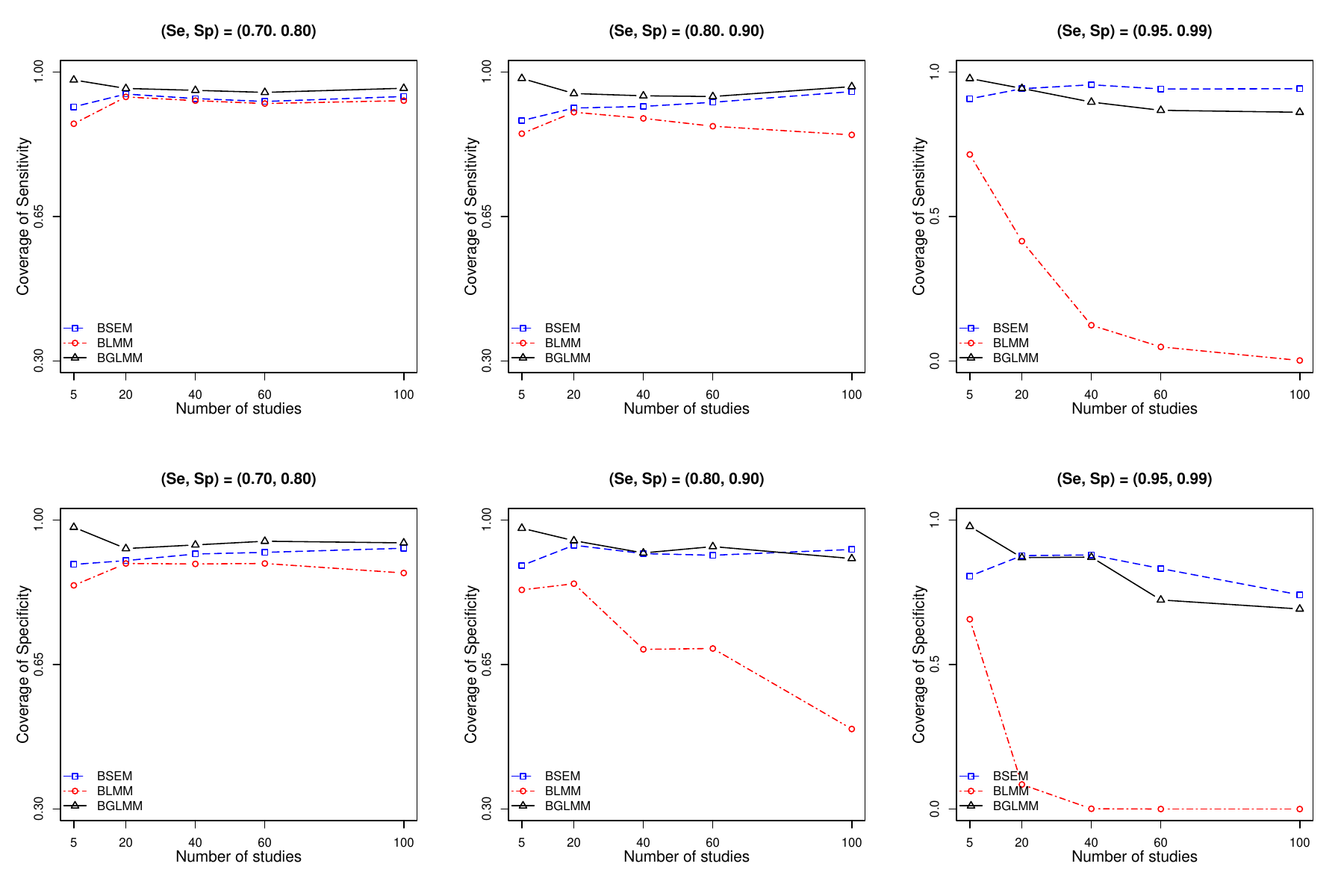}
        \caption{Coverage probabilities of pooled sensitivity and specificity for different combinations of true sensitivity, true specificity, and number of studies when $\sigma_1^2=\sigma_2^2=0.80$, $\sigma_{12}=-0.50$, $n_1=40$, $n_2=50$ and random effects were generated from the bivariate normal distribution.}
        \label{Fig7}
    \end{figure}
\end{center}

Figure \ref{Fig7} depicts the empirical 95\% CPs for the true sensitivity and specificity generated by the three models. In the low-diagnostic accuracy setting, the models had comparable CPs, although the standard BGLMM had CPs relatively closer to the nominal 95\% than the BSEM and BLMM did. However, the standard BLMM yielded poorer CPs that approached zero as the number of studies increased in moderate-to-high diagnostic accuracy settings. Under the same settings, the proposed BSEM achieved CPs comparable to or better than those of the standard BGLMM, particularly in high-diagnostic-accuracy settings. The models had comparable CPs for meta-analyses with large sample sizes within the study, especially when the number of studies was at least 20. However, the standard BLMM had lower CPs for specificity, which were farther from the nominal values than those of the BGLMM and BSEM (Appendix Figures E.3, E.6, and E.9). Changing the random effects distribution to the bivariate $t$ or Laplace distribution did not affect the trend of the CP results seen for the bivariate normal distribution scenarios (Appendix Figures E.10 to E.27).

\begin{center}
    \begin{figure}[h]
        \centering
        \includegraphics[width = \textwidth]{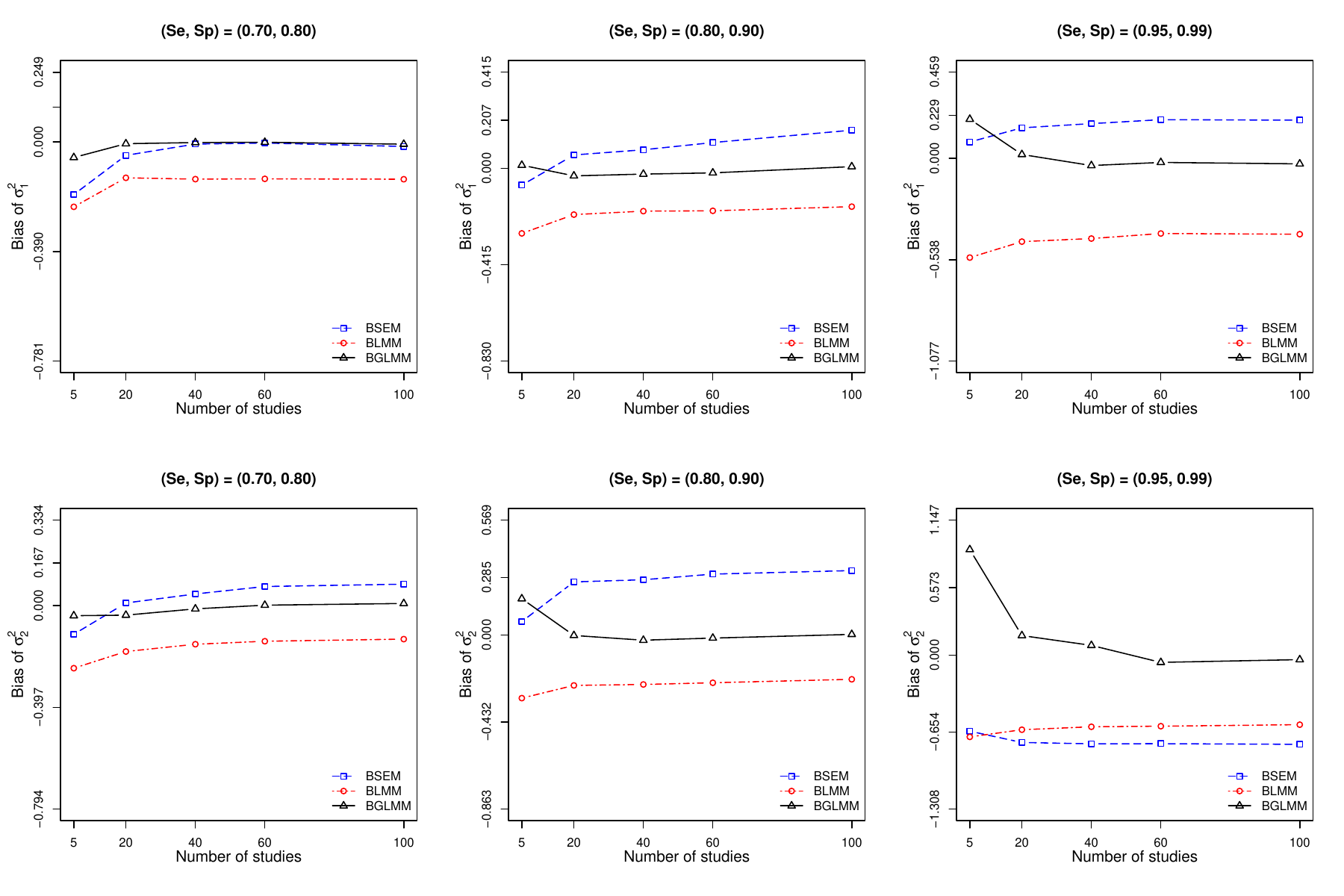}
        \caption{Bias of between-study variances for different combinations of true sensitivity, true specificity, and number of studies when $\sigma_1^2=\sigma_2^2=0.80$, $\sigma_{12}=-0.50$, $n_1=40$, $n_2=50$ and random effects were generated from the bivariate normal distribution.}
        \label{Fig8}
    \end{figure}
\end{center}

Figure \ref{Fig8} shows the bias in between-study variances for the logit-transformed sensitivity and specificity. The standard BLMM severely underestimated both variances regardless of the true diagnostic accuracy values. The proposed BSEM approach had bias values closer to zero and BGLMM values in low-diagnostic-accuracy settings. However, the standard BGLMM had a lower bias than the other two models in the moderate-to-high diagnostic-accuracy settings. The models produced comparable, lower bias values that approached zero as the number of participants in a study increased (Appendix Figures F.3, F.6 and F.9). Consistent results were obtained when random effects were generated from the bivariate $t$ and Laplace distributions (Appendix Figures F.10 to F.27).

\begin{center}
    \begin{figure}[h]
        \centering
        \includegraphics[width = \textwidth]{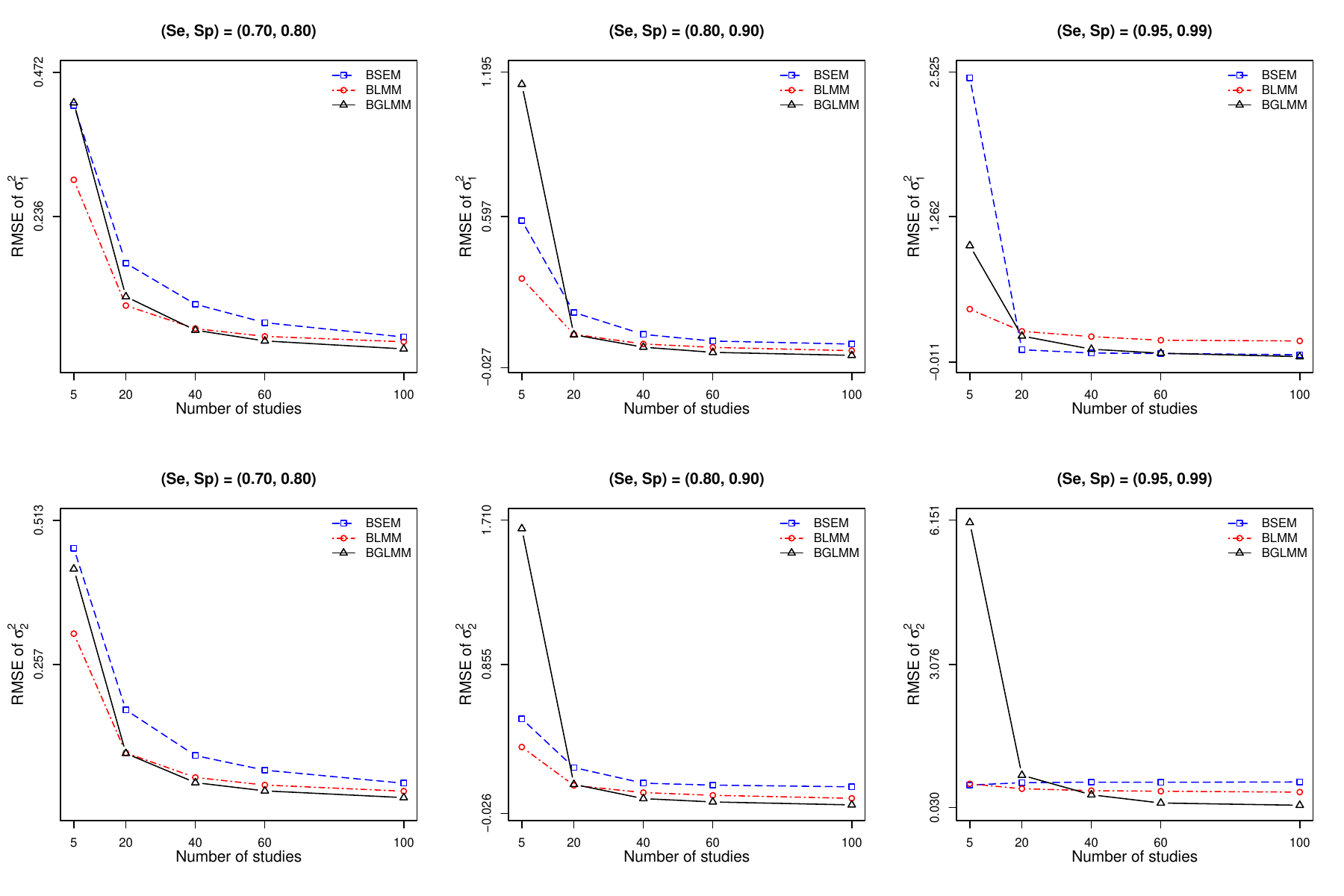}
        \caption{RMSE of between-study variances for different combinations of true sensitivity, true specificity, and number of studies when $\sigma_1^2=\sigma_2^2=0.80$, $\sigma_{12}=-0.50$, $n_1=40$, $n_2=50$ and random effects were generated from the bivariate normal distribution.}
        \label{Fig9}
    \end{figure}
\end{center}

In terms of RMSE for the between-study variances (Figure \ref{Fig9}), all models' RMSE values decreased as the number of studies increased, as expected. The standard models, BGLMM and BLMM, produced comparable, marginally lower RMSEs than the proposed BSEM in low-to-moderate diagnostic-accuracy settings. However, the BSEM had RMSE values comparable to those of the standard models in high-diagnostic accuracy settings. All three models had similar RMSEs that approached zero faster as the study-specific number of participants increased (Appendix Figures G.3, G.6, and G.9). These results were consistent when random effects were generated from the bivariate $t$ and Laplace distributions (Appendix Figures G.10 to G.27).

\begin{center}
    \begin{figure}[h]
        \centering
        \includegraphics[width = \textwidth]{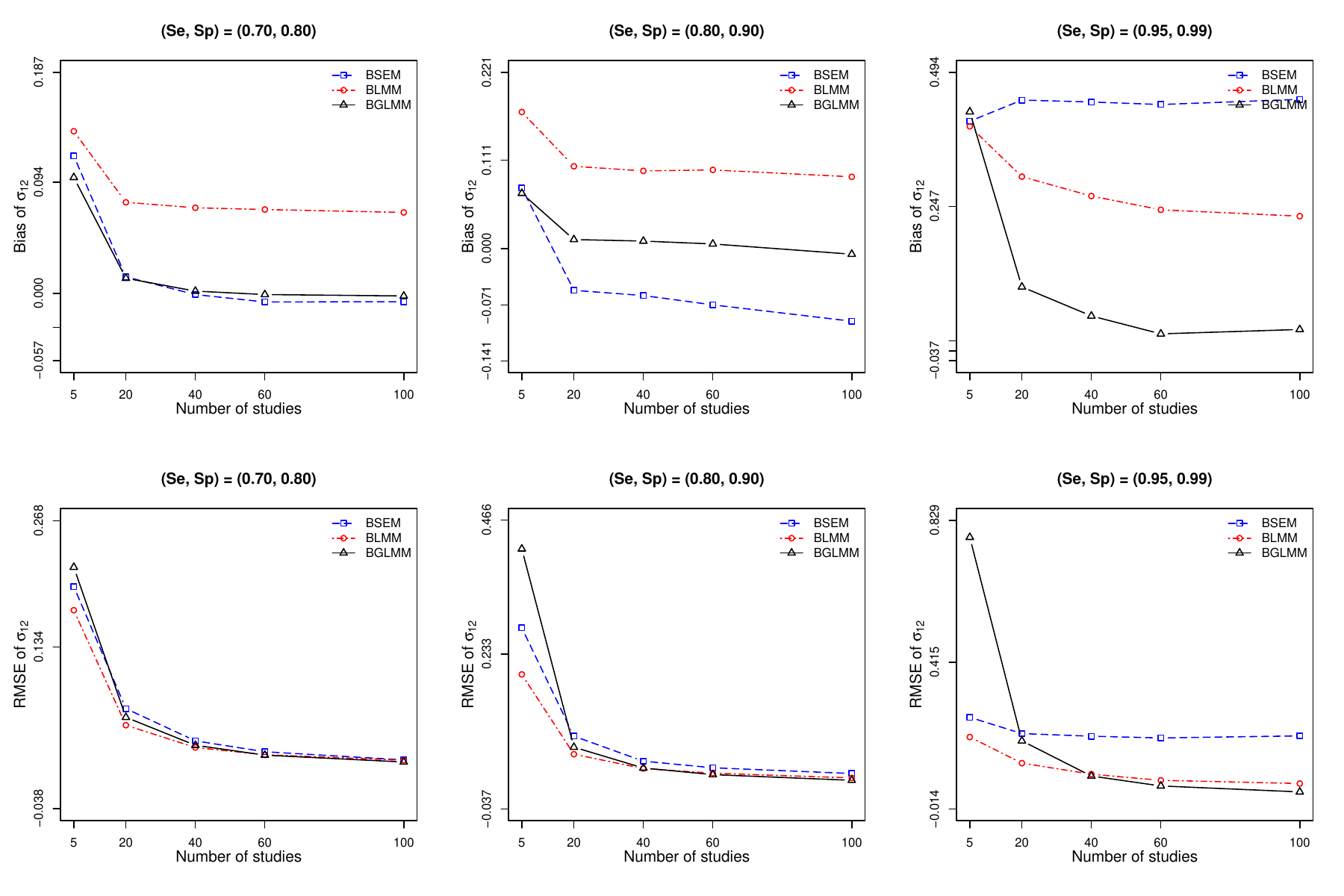}
        \caption{Bias (top row) and RMSE (bottom row) of between-study covariance for different combinations of true sensitivity, true specificity, and number of studies when $\sigma_1^2=\sigma_2^2=0.80$, $\sigma_{12}=-0.50$, $n_1=40$, $n_2=50$ and random effects were generated from the bivariate normal distribution.}
        \label{Fig10}
    \end{figure}
\end{center}

Figure \ref{Fig10} depicts the bias and RMSE of the between-study covariance parameter. The standard BLMM consistently overestimated $\sigma_{12}$ irrespective of the different parameters in our simulation study. The proposed BSEM had a comparable or better bias than the BGLMM in low-to-moderate diagnostic-accuracy settings, whereas the opposite was true in the high diagnostic-accuracy setting. All models had comparable RMSE values, except that those of the BSEM were marginally higher for some of the high diagnostic-accuracy settings. Increasing the number of participants per study made the bias and RMSE values more similar across models and moved them toward zero more quickly as the number of studies increased (Appendix Figures H.3, H.6, and H.9). Consistent results were obtained when changing the random effects distribution to bivariate $t$ or Laplace (Appendix Figures H.10 to H.27).

Finally, Table~\ref {Tab2} summarizes the convergence rates of the three models across the simulation scenarios. Overall, all three models showed relatively stable convergence behaviour across the three random-effects distributions, although convergence rates varied between models. The BLMM achieved a 100\% convergence rate under all three distributions. The proposed BSEM converged in 92.2\% of simulation replicates when the random effects were generated from a bivariate normal distribution, compared to 100.0\% for the BLMM and 87.5\% for the BGLMM. When random effects were generated from a bivariate $t$ distribution, the corresponding convergence rates were 90.8\%, 100.0\% and 84.4\%, respectively. Under the bivariate Laplace distribution, the convergence rates were 91.4\%, 100.0\% and 86.3\%, respectively. Thus, BSEM consistently achieved higher convergence rates than BGLMM across the three random-effects distributions and maintained rates above 90\% in all cases. The BSEM convergence rate was also relatively stable in all three distributions, ranging from 90.8\% to 92.2\%, while the BGLMM had the lowest convergence, with rates ranging from 84.4\% to 87.5\%.

\begin{table}[ht]
\centering
\caption{Convergence rates, in percent, of the models under three random-effects distributions}
\label{Tab2}
\begin{tabular}{@{}lccc@{}}
\toprule
\textbf{Random-effects distribution} &
\textbf{BSEM} &
\textbf{BLMM} &
\textbf{BGLMM} \\
\midrule
Bivariate normal  & 92.2 & 100.0 & 87.5 \\
Bivariate $t$     & 90.8 & 100.0 & 84.4 \\
Bivariate Laplace & 91.4 & 100.0 & 86.3 \\
\bottomrule
\end{tabular}
\end{table}

\section{Illustrative examples}
\label{S5}
In this Section, we revisit the two real-life datasets introduced in Section \ref{S2}. We reanalyze them using the standard and proposed approaches.

\subsection{IAQ data}
Table \ref{Tab3} presents the estimates obtained from the BGLMM, BLMM, and proposed BSEM approach. For the IAQ dataset, the three models produced very similar estimates of pooled sensitivity and specificity, although those of the BLMM estimates were slightly lower (by about 1\%) than those of the other two models. The direction of these differences aligned with our simulation results, where the BLMM estimates were generally negatively biased relative to the known data-generating values. The corresponding 95\% CIs overlapped substantially, although the BLMM had the narrowest CI width, which also aligns with our simulations for settings with meta-analysis of size $k \leq 20$.

Regarding the heterogeneity parameters between studies, the estimates showed somewhat greater differences across models. The BLMM model tended to produce smaller estimates of between-study variance than the BSEM and BGLMM. The direction of these differences was consistent with the simulation results, in which the BLMM generally showed greater bias and tended to underestimate between-study variance. The estimated correlations between the random effects were negative and moderate for all three models, indicating a moderate trade-off between sensitivity and specificity in the studies (i.e., as sensitivity increases, specificity tends to decrease moderately). Although the non-diseased group appeared to have a relatively higher degree of absolute heterogeneity ($\sigma_2^2=2.388$) compared to the diseased group ($\sigma_1^2=1.755$), the LRT for the equality of heterogeneity variances based on the fitted BSEM gave a $p$-value of $0.431$, which does not provide evidence against the null hypothesis of equal variances at the 5\% significance level.

Appendix Figure A.1 presents 95\% confidence regions\cite{Cheung2013MultivariateSEM} for the IAQ dataset after fitting the proposed BSEM approach. The large green ellipse describes the heterogeneity between studies and shows where we expect the true sensitivity and specificity of a new future study to fall 95\% of the time. The downward tilt of both the red summary ellipse and the giant green random-effects ellipse confirms a classic diagnostic threshold effect. In IAQ studies, as the sensitivity increases, specificity decreases, strongly indicating that different primary studies used different diagnostic cut-off points to define a positive test result. The green arrows show that while the pooled specificity is highly stable with a narrow CI, the sensitivity varies greatly between studies.

\begin{table}[ht]
\caption{Estimates of pooled sensitivity and specificity along with their 95\% CIs, estimates of between-study heterogeneity parameters, and likelihood ratio test for the random-effects components for the IAQ\cite{Patrick1994ValiditySelfReportedSmoking} and DVT\cite{Kearon1998NoninvasiveDiagnosis} datasets}
\label{Tab3}
\centering
\begin{tabular}{@{}lllllllll@{}}
  \toprule
\textbf{Model} & \textbf{Sensitivity (95\% CI)} & \textbf{Specificity (95\% CI)} & $\sigma_1^2$ & $\sigma_2^2$ & $\sigma_{12}$ & $\rho$ & \textbf{LRT ($p$-value)} \\ 
  \midrule
   &   & \textbf{IAQ dataset} & & & & &   \\ \midrule 
BGLMM & 0.960 (0.931, 0.977) & 0.932 (0.868, 0.966) & 1.326 & 2.203 & -1.037 & -0.607 &  \\ 
BLMM & 0.955 (0.930, 0.972) & 0.927 (0.871, 0.960) & 0.956 & 1.906 & -0.838 & -0.621 &  \\ 
BSEM & 0.965 (0.935, 0.981) & 0.934 (0.874, 0.967) & 1.755 & 2.388 & -1.194 & -0.583 & 0.620 (0.431) \\ \midrule
   &   & \textbf{DVT dataset} & & & & &   \\ \midrule 
BGLMM & 0.764 (0.650, 0.850) & 0.965 (0.943, 0.979) & 1.987 & 1.474 & -0.259 & -0.151 &  \\ 
BLMM & 0.752 (0.650, 0.832) & 0.955 (0.932, 0.970) & 1.617 & 0.883 & -0.338 & -0.283 &  \\ 
BSEM & 0.765 (0.654, 0.848) & 0.970 (0.953, 0.981) & 1.931 & 1.446 & -0.290 & -0.173 & 0.529 (0.467) \\ 
  \bottomrule
\end{tabular}
\end{table}

\subsection{DVT data}
For the DVT dataset, the models again provided similar estimates of pooled sensitivity and specificity with overlapping 95\% CIs (Table \ref{Tab3}). Moreover, BLMM again yielded lower estimates of sensitivity and specificity, with a narrower sensitivity CI. The direction of the difference between the models was consistent with the simulation results for meta-analyses of size $k=30$ and moderate sensitivity and specificity (Figures \ref{Fig4} and \ref{Fig6}, left panel). Furthermore, although all models estimated very high specificity, the BSEM produced a narrower CI, consistent with the pattern observed in the simulations for highly specific meta-analyses (Figure \ref{Fig6}, right panel).

Again, estimates of the between-study heterogeneity parameters showed variation between models. The BLMM produced smaller estimates of the between-study variance than the BGLMM and BSEM. The direction of this difference was consistent with the simulation results, in which the BLMM tended to underestimate the between-study variance. All three models yielded negative but mild threshold-driven correlations between sensitivity and specificity, although the magnitude of the correlation differed between models. Despite the diseased group appearing more heterogeneous than the non-diseased group, the LRT for equality of variances yielded a $p$-value of $0.467$, providing no evidence against the null hypothesis at the 5\% level of significance.

Appendix Figure A.2 shows the confidence regions for the DVT data. It shows that venous ultrasonography is very specific and generally produces fewer false positives. The large variation between studies in the DVT data does not seem driven by a simple threshold trade-off, since the green random-effects ellipse is oriented almost circularly rather than slanting downward, reinforcing the estimated mild negative correlation. Thus, heterogeneity between studies may reflect differences in participant populations, clinical settings, or reference-standard quality. To assess this, we fitted meta-regression models, adjusting for study-level participant characteristics (asymptomatic vs symptomatic). The meta-regression results are presented in Appendix Table A.3. As expected, all models yielded very low baseline sensitivity [$e^{\widehat{\beta}_{0_{Se}}}/(1+e^{\widehat{\beta}_{0_{Se}}})$] and very high baseline specificity [$e^{\widehat{\beta}_{0_{Sp}}}/(1+e^{\widehat{\beta}_{0_{Sp}}})$] estimates that were statistically significant for asymptomatic participants. In addition, the models produced significantly higher sensitivity for symptomatic participants [$e^{\widehat{\beta}_{0_{Se}}+\widehat{\beta}_{1_{Se}}}/(1+e^{\widehat{\beta}_{0_{Se}}+\widehat{\beta}_{1_{Se}}})$] than for asymptomatic participants. The estimated specificities for symptomatic participants [$e^{\widehat{\beta}_{0_{Sp}}+\widehat{\beta}_{1_{Sp}}}/(1+e^{\widehat{\beta}_{0_{Sp}}+\widehat{\beta}_{1_{Sp}}})$] were slightly lower than for asymptomatic participants, although the differences were not statistically significant. To evaluate how much of the unexplained variation between studies was accounted for by adjusting for participant symptoms, we calculated the percentage change in the random effects parameters from the baseline models in Table \ref{Tab3} to the meta-regression models in Appendix Table A.3. The meta-regression reveals a massive drop of 67\% to 73\% in the sensitivity variance $(\sigma_1^2)$ across all three models. This reaffirms that participant symptoms (asymptomatic vs. symptomatic) were the primary drivers of the heterogeneity in test sensitivity between studies. In contrast, the variance in specificity $(\sigma_2^2)$ remains almost completely unchanged, again confirming the meta-regression findings that this participant characteristic does not explain the variation in specificity.

\section{Discussion and Conclusion}
\label{S6}

We evaluated the performance of the proposed BSEM against two widely used standard approaches for ADMA of DTA studies: the BLMM of Reitsma et al. (2005)\cite{Reitsma2005BivariateAnalysis} and the BGLMM of Chu and Cole (2006)\cite{Chu2006BivariateMetaAnalysis} through an extensive simulation study. The relative performance of the three approaches depended on the diagnostic-accuracy level, the number of studies, and the sample size per study. Overall, the proposed BSEM was comparable to the BGLMM or outperformed the BLMM in estimating pooled sensitivity and specificity and their associated uncertainties. In addition, the proposed BSEM has flexible characteristics that unify many extensions of standard models, which we discuss further below.

An important finding was the strong influence of the within-study number of participants on the performance of all three methods. When the numbers of diseased and non-diseased participants were increased, the differences among the models largely disappeared. Bias and RMSE approached zero, CIs became increasingly similar, and empirical CPs became more comparable, particularly when the number of studies was also large. Thus, the principal differences between the methods emerged in settings in which both the number of studies and the amount of information within the individual studies were limited. This is an important practical consideration because DTA meta-analytic data frequently involve relatively small numbers of studies and heterogeneous studies.\cite{Mathes2018SmallNumberBinary, Plana2021HeterogeneityDiagnostic} Therefore, in such settings, the choice of model could have a meaningful effect on both point estimation and inference about pooled sensitivity and specificity.

Our proposed BSEM is particularly advantageous in DTA meta-analysis settings characterized by small within-study sample sizes, relatively few studies, and high diagnostic accuracy. Under these conditions, the BSEM generally provided less-biased pooled sensitivity and specificity estimates than the BLMM, competitive RMSEs, and comparable or better empirical coverage than the BGLMM. Moreover, BSEM demonstrated generally stable computational performance, with convergence rates exceeding 90\% across all three random-effects distributions. The simulation results also showed that the apparent precision of the BLMM can be misleading since, although its CIs were often narrow, its empirical CP was substantially below the nominal level in several important scenarios. In addition, model choice became less consequential when meta-analyses included many studies with large within-study sample sizes because, as information increased, differences among the three methods decreased. Furthermore, changing the distribution used to generate the random effects from bivariate normal to bivariate $t$ or the Laplace distribution did not significantly alter the main conclusions about the pooled sensitivity and specificity for all three models. The models' relative performance remained similar, and we observed the same general patterns in bias, RMSE, CI width, and empirical CP. Finally, applying the models to two empirical DTA datasets demonstrated the approach's utility for practitioners and showed differences between the models that generally aligned with patterns observed in the simulation study.

The strengths of this manuscript include introducing an SEM-based approach and inferential methods for meta-analyzing DTA studies, evaluating the proposed BSEM against well-established approaches through extensive simulation studies, and demonstrating the practical utility of the proposed method using real-world datasets. However, the current study also has limitations that present opportunities to further develop the proposed BSEM framework. We considered settings that compare an index test with a perfect reference standard, although imperfect reference standards are common in many DTA applications. We also compared a single index test against a reference standard, although real-world scenarios may involve comparing multiple DTA tests against the same reference standard. Our simulation study also did not account for meta-analyses with study-level covariates to fit a meta-regression model or for DTA studies with partial verification bias due to not all participants receiving the reference standard or missing test outcomes.\cite{Collins2014DiagnosticTestAccuracy}

Because the proposed BSEM is flexible, future studies can extend it to address the limitations mentioned above. An SEM framework that unifies LCM and SEM-based meta-analytic modelling can handle imperfect reference standards by interpreting sensitivity and specificity as conditional response probabilities when constructing the conditional measurement model, modelling study-specific parameters as random effects, and constructing a full hierarchical likelihood by combining the measurement and structural components. However, it should be noted that in this extension, the SEM-based meta-analytic parameters will not be identifiable without additional assumptions, such as the need for at least three diagnostic tests, parameter constraints (e.g., fixing sensitivity, specificity, or assuming known prevalence), or posing informative priors if the model is fit in a Bayesian context. The proposed BSEM also extends naturally to the setting where multiple diagnostic tests are evaluated simultaneously.\cite{Pereira2012GeneralLatentClass} In this extension of the BSEM, multiple observed indicators (index tests) will load on either latent accuracy parameters (in models with perfect reference standards) or on latent disease status in imperfect reference standard models. This multivariate extension would allow for joint estimation of multiple sensitivities and specificities, modelling of correlation between tests, and evaluation of index test combinations. As demonstrated in our real-life data examples, meta-regression within the BSEM framework is also possible, as covariates can be incorporated into SEM-based DTA models naturally. For example, study-level covariates such as study design or patient characteristics can be included as predictors of latent sensitivity and specificity or of disease prevalence; test-level covariates can influence measurement parameters; and study-level moderators can help explain between-study heterogeneity due to differences in thresholds, population characteristics, or test execution. Furthermore, the proposed BSEM approach is again suitable for modelling DTA studies with partial verification bias, because missingness can be handled via a full-information likelihood and disease status can be modelled as latent when verification is incomplete. Finally, developing an extension of the proposed BSEM approach for meta-analysis of individual-participant data from DTA studies is a worthwhile future study to evaluate the framework's performance in this gold-standard meta-analysis setting.

In conclusion, our simulation study and real-world data applications provided evidence that the proposed BSEM is a flexible alternative to existing bivariate approaches for DTA meta-analysis. Its principal advantage is that BSEM provided stable, better estimates across a broad range of practically relevant scenarios, with particularly favourable pooled sensitivity and specificity when the meta-analysis size was small and the index test had high diagnostic accuracy. The results therefore supported using BSEM as an alternative modelling approach for ADMA of DTA studies, although the relative advantages of the standard methods depended on the meta-analytic data size, number of participants within the study, and the specific parameter being estimated by the models.

\paragraph{Acknowledgments.}
The author is grateful to the computing facilities provided by the Faculty of Mathematics at the University of Waterloo.

\paragraph{Funding Statement.}
This research was supported by funding from the Natural Sciences and Engineering Research Council of Canada (NSERC), grant number RGPIN-2024-04652.

\bibliographystyle{unsrtnat}
\bibliography{references}  






\end{document}